\documentclass{emulateapj}
\usepackage{graphicx}
\usepackage{natbib}
\usepackage{bm}
\usepackage{verbatim}
\usepackage[usenames]{color}

\usepackage[fleqn]{amsmath}
\usepackage{amssymb}
\usepackage{color}

\newcommand\icarus{Icarus}

\newcommand{\ob}{\Omega_\mathrm{b}}

\newcommand{\ab}{a_\mathrm{b}}
\newcommand{\eb}{e_\mathrm{b}}

\newcommand{\msun}{\mathrm{M_\odot}}
\newcommand{\rinf}{R_\mathrm{inf}}

\newcommand{\aP}{a_\mathrm{p}}
\newcommand{\ep}{e_\mathrm{p}}
\newcommand{\ef}{e_\mathrm{f}}
\newcommand{\eff}{e_\mathrm{ff}}

\begin{document}

\title{How not to build Tatooine: the difficulty of in situ formation of circumbinary planets Kepler 16b, Kepler 34b and Kepler 35b}
\author{
Sijme-Jan Paardekooper\altaffilmark{1},
Zo\"e M. Leinhardt\altaffilmark{2},
Philippe Th\'ebault\altaffilmark{3},
Cl\'ement Baruteau\altaffilmark{1}
}

\altaffiltext{1}{DAMTP, Wilberforce Road, Cambridge CB3 0WA, United Kingdom; S.Paardekooper@damtp.cam.ac.uk}
\altaffiltext{2}{School of Physics, University of Bristol, H.H. Wills Physics Laboratory, Tyndall Avenue, Bristol BS8 1TL, U.K.}
\altaffiltext{3}{Observatoire de Paris, F-92195 Meudon Principal Cedex, France}

\begin{abstract}
We study planetesimal evolution in circumbinary disks, focusing on the three systems Kepler 16, 34 and 35 where planets have been discovered recently. We show that for circumbinary planetesimals, in addition to secular forcing, eccentricities evolve on a dynamical timescale, which leads to orbital crossings even in the presence of gas drag. This makes the current locations of the circumbinary Kepler planets hostile to planetesimal accretion. We then present results from simulations including planetesimal formation and dust accretion, and show that even in the most favourable case of 100\% efficient dust accretion, in situ growth starting from planetesimals smaller than $\sim{10}~\mathrm{km}$ is difficult for Kepler 16b, Kepler 34b and Kepler 35b. These planets were likely assembled further out in the disk, and migrated inward to their current location. 
\end{abstract}

\keywords{planets and satellites: formation}
\maketitle

\section{Introduction}

In the last decade planets have been found in very perturbed systems such as close binary star systems. The first of these planets to be discovered were orbiting the primary star \citep{queloz00,hatzes03,zucker04}, but the latest additions to the family, after promising results using stellar eclipse timings \citep{lee09}, involve planets in circumbinary orbits: Kepler 16 \citep{doyle11} and Kepler 34 and 35 \citep{welsh12}. The parameters of these new planets are summarised in Table \ref{tab1}.

The existence of planets in these systems baffles planet formation theory. A crucial step in the process of building a planet, namely growing gravitationally bound protoplanets from km-sized planetesimals, can be hindered or stopped in these perturbed environments for planetesimals on circumprimary orbits \citep{marzari00,thebault06,paard08,thebault11}. The coupling between gravitational perturbations of the companion star and gas drag stirs up the eccentricities of planetesimals, which leads to high encounter velocities. This makes accretion towards larger bodies difficult. Similar problems haunt planetesimals on circumbinary orbits\citep{moriwaki04,scholl07,marzari08,meschiari12}. 

Above studies focused on gravitational dynamics and gas drag only. In this work, we investigate the effect of collisions on the evolution of the system. \cite{collision} showed that in a system with high-speed collisions, it is necessary to keep track of collision outcomes. Notably, if collisions are mostly destructive, any surviving planetesimals are embedded in a sea of small debris. If they can pick up some of this debris, planetesimals can grow despite the hostile environment \citep{collision,xie10}. In this Letter, we aim to explore this possibility in the newly found planet-harbouring systems of Kepler 16, 34 and 35. 

We begin in Section \ref{secEcc} by reviewing the eccentricity evolution of planetesimals in circumbinary orbits. We discuss the model in Section \ref{secMod}, present the results in Section \ref{secRes}, and conclude in Section \ref{secDisc}.

\begin{deluxetable}{rrrr} 
\tablecolumns{4} 
\tablewidth{0pc} 
\tablecaption{Binary and planet parameters} 
\tablehead{ 
\colhead{ } & \colhead{Kepler 16$^a$} & \colhead{Kepler 34$^b$}   & \colhead{Kepler 35$^b$} 
}
\startdata 
$M_A/\msun$ & 0.69 & 1.0 & 0.89\\
$M_B/\msun$ & 0.20 & 1.0 & 0.81\\
$\ab/$AU & 0.22 & 0.23 & 0.18\\
$\eb$ & 0.16 & 0.52 & 0.14\\
$M_\mathrm{p}/M_\mathrm{J}$ & 0.33 & 0.22 & 0.13\\
$\aP/\ab$ & 3.2 & 4.7 & 3.4\\
$\ep$ & 0.0069 & 0.18 & 0.042
\enddata 
\tablenotetext{a}{\cite{doyle11}}
\tablenotetext{b}{\cite{welsh12}}
\label{tab1}
\end{deluxetable} 

\section{Secular and non-secular eccentricity evolution}
\label{secEcc}

Consider the dynamics of massless particles at semi-major axis $a$ and period $2\pi/\Omega$ around a binary with masses $M_A$ and $M_B$, total mass $M_*=M_A+M_B$, orbital period $2\pi/\ob$, eccentricity $\eb$ and semi-major axis $\ab$. Secular perturbation theory gives an evolution equation for the complex eccentricity $E$ of planetesimals in the gas-free case \citep{moriwaki04}:
\begin{equation}
\frac{1}{\ob^2}\frac{d^2E}{d\tau^2}=E_\mathrm{f}-E,
\label{eqEcc}
\end{equation}
where $\tau$ is the secular timescale and $E_\mathrm{f}=e_\mathrm{f}\exp(i\varpi_\mathrm{b}) $ is the complex forcing, with $\varpi_\mathrm{b}$ the longitude of periastron of the binary orbit and
\begin{equation}
e_\mathrm{f}=\frac{5}{4}\frac{M_A-M_B}{M_*}\frac{\ab}{a}\eb\frac{1+3\eb^2/4}{1+3\eb^2/2}
\end{equation} 
the forced eccentricity \citep{moriwaki04}. The secular timescale is given by \citep{moriwaki04}:
\begin{equation}
\Omega_\mathrm{b}\tau=\frac{4}{3}\frac{M_*^2}{M_AM_B}\left(\frac{a}{\ab}\right)^{7/2}\frac{1}{1+\frac{3}{2}\eb^2}.
\end{equation}
Equation (\ref{eqEcc}) describes an oscillation around the forced eccentricity with an amplitude given by $|E(t=0)-E_\mathrm{f}|$. The period of oscillation is the secular timescale, which is longer than the dynamical timescale for $a\gg\ab$. While initially, planetesimal orbits are phased, the spatial frequency of the oscillations increases with time, so that eventually orbital crossings occur leading to high encounter velocities \citep{thebault06}. In the presence of a gas disk and associated drag forces, these oscillations are damped, and size-dependent equilibrium orbits exist \citep{paard08}. Even if orbital crossings can be prevented, the size-dependence of equilibrium orbits leads to high encounter velocities between bodies of different size. This is called differential orbital phasing \citep{thebault06}. For equal-mass binaries, such as Kepler 34 and 35, $\ef\sim0$. While this is favourable for accretion, it is not the whole story. 

Secular perturbation theory is valid on timescales longer than both the binary period and the local orbital timescale. In addition, planetesimal eccentricities evolve on a shorter timescale, even in the case $\eb=0$, where secular effects are absent \citep{moriwaki04}. This short timescale evolution can be obtained by averaging the disturbing function over the mean longitude of the binary orbit only, and expanding terms up to second order in both $\eb$ and planetesimal eccentricity $e$. The eccentricity can then be shown to oscillate, on a local orbital timescale, around an eccentricity
\begin{equation}
\eff=\frac{3}{4}\frac{M_AM_B}{M_*^2}\left(\frac{\ab}{a}\right)^2\sqrt{1+\frac{34}{3}\eb^2},
\end{equation}
where the subscript ff indicates a fast timescale. Note that, unlike secular oscillations, the period is independent of $\eb$, and that $\eff$ falls off faster with distance than $\ef$. In the case of almost equal mass binaries (like Kepler 34 and 35), $\ef$ will be very small, while $\eff$ can be significant. Since $e$ now evolves on a local dynamical timescale, orbital crossings will occur after only a few local orbits. Gas drag is unable to damp these fast oscillations, because it acts on longer timescales than a dynamical timescale for km-sized objects and realistic gas densities. Orbit crossings cannot be prevented. Typical encounter velocities will be $\sim\eff{a}\Omega$ ($250~\mathrm{m/s}$ in Kepler 16, and $500~\mathrm{m/s}$ in Kepler 34 and 35 at the locations of the planets). Despite the lack of secular forcing, this makes the systems Kepler 34 and 35 hostile to accretion at the current planet positions. For Kepler 16, encounter velocities around $250~\mathrm{m/s}$ can lead to accretion only when planetesimals have reached sizes of $50~\mathrm{km}$ or larger \citep{thebault06}. At twice the current semi-major axis of Kepler 16b, encounter velocities go down a factor of 4, making accretion possible at that location if planetesimals can somehow grow to $\ge10~\mathrm{km}$  \citep[also seen by][]{meschiari12}.
  
\section{Model ingredients}
\label{secMod}

In our simulations we consider a system of two stars with a coplanar circumbinary disk. The gas component of the disk is assumed to be circular and orbiting the binary centre of mass. The solid component of the disk consists of planetesimals $\ge1$ km, which we model as particles, and small dust, on the same orbits as the gas. Planetesimals can form from small dust, accrete small dust on their surface, and be returned to dust in catastrophic collisions. Below, we explain the planetesimal evolution model in more detail. 

\subsection{Two-dimensional approximation}
\label{sec2D}
As in \cite{collision}, we restrict ourselves to planetesimal orbits lying in the orbital plane of the binary. Preliminary calculations allowing for inclination of planetesimal orbits show no qualitative differences. In addition, allowing for three-dimensional (3D) motions requires an impractical increase in CPU time. Moreover, we expect collision velocities induced by fast or secular forced eccentricities to be much higher than the escape velocity from the surface of the planetesimal (catastrophic collisions), which, as we explain below, makes the motions essentially two-dimensional (2D). However, confining all orbits to a single plane will underestimate the true 3D collision timescale \citep{collision}, strongly affecting the evolution of the system, which is sensitive to the balance between dynamical and collisional effects. In a strongly perturbed system, where almost every collision is destructive, it is difficult to estimate the 3D collision timescale since it is unclear what the inclination distribution of the fragments will be. For catastrophic collisions, \cite{leinhardt12} showed that the maximum velocities obtained by the largest fragments are comparable to the escape speed of the combined projectile-target mass. Therefore, in the regime of high-speed collisions expected in a close binary system the planetesimal disk is expected to stay approximately as thin as in the unperturbed case (assuming the largest collision remnants trace the majority of the mass). Thus, the thickness of the disk is set by the escape velocity of the planetesimals. In 2D, we can vary the collision timescale by changing the total mass in solids (results do not depend sensitively on this mass). Note that the thin disk approximation does not necessarily apply to the small debris. If the dust disk thickens because of collisions both dust accretion and planetesimal formation will be effected. 

\subsection{Planetesimal formation and orbital phasing}
Planetesimal dynamics in binary systems is strongly affected by the timescale on which planetesimals are formed. If planetesimals form fast compared to the timescale of eccentricity forcing, and in a single burst, the interaction with the binary stars will keep their orbits in phase and collision velocities low \citep{hep78}. However, size-dependent gas drag introduces \emph{differential} orbital phasing, which leads to high collision velocities between bodies of different size, making planetesimal accretion difficult \citep{thebault06}. Eccentricity forcing towards $\eff$ occurs on a dynamical timescale. On such short timescales orbital crossings can not be prevented by gas drag. In addition, if planetesimals form continuously, their orbits will be out of phase from the start. The local collision velocity is estimated as, $v_\mathrm{col}\approx\bar{e}a\Omega$, where $\bar{e}$ is a typical eccentricity.  

A reasonable timescale for the formation of an individual planetesimal is $10^4$ years \citep{lissauer93}. This places a lower limit on the timescale for dust to be converted into planetesimals. \cite{chambers10} found that the latter timescale could vary by several orders of magnitude depending on the local conditions in the disk. In the simulations presented here, as in \cite{collision}, planetesimals form continuously with half of the total (local) dust mass converted into planetesimals in $10^5$ local orbits. Planetesimals that wander off the computational domain are added to the innermost or outermost dust bin, thereby allowing them to be recycled.

\subsection{Collisions}
\label{secColTime}
In previous, purely dynamical codes, planetesimals eventually reached steady-state orbits imposed by gravitational perturbations and gas drag \citep[e.g.][]{marzari00,thebault06,scholl07,xie09,meschiari12}. In the work presented here this is not necessarily the case, because collisions can affect the size of planetesimals on their way to their steady-state orbits.  

If we assume that the thickness of the planetesimal disk is not strongly affected by high-speed collisions (Sec. \ref{sec2D}) and a typical collision speed $v_\mathrm{col}=\bar{e}a\Omega$, we can estimate the collision timescale in a 3D disk. Consider a population of planetesimals of size $R$, mass $M$, and number density $n$. The collision timescale is:
\begin{eqnarray}
\Omega\tau_\mathrm{c}=\frac{\Omega}{\pi{R}^2nv_\mathrm{col}}=\nonumber\\
\frac{0.5}{\bar{e}}\left(\frac{a}{\mathrm{AU}}\right)^{\frac{1}{2}}\left(\frac{R}{\mathrm{km}}\right)^2\frac{17 \mathrm{~g/cm^{2}}}{\Sigma}\left(\frac{\rho_p}{3\mathrm{~g/cm^{3}}}\right)^{\frac{3}{2}}\left(\frac{M_\odot}{M_*}\right)^{\frac{1}{2}},
\end{eqnarray}
where we have used $n=\Sigma/(2\Delta{z}M)$, with $\Sigma$ the surface density of planetesimals and a thickness $\Delta{z}$ set by the escape velocity, $\Delta{z}=v_\mathrm{esc}/(2a\Omega)$. A similar expression can be derived in the 2D approximation, yielding a ratio:
\begin{equation}
\frac{\tau_\mathrm{c,3D}}{\tau_\mathrm{c,2D}}=2073\left(\frac{a}{\mathrm{AU}}\right)^{3/2},
\end{equation}
which depends only on $a$. To get realistic collision timescales in our 2D approximation, we can tune the $\tau_\mathrm{2D}$ value to $\tau_\mathrm{3D}$ by artificially reducing the disk mass by a factor 2000. This setting is, to first order, an acceptable one because we have found that the qualitative outcome of the simulations does not depend sensitively on the disk mass.  
  
Having chosen a disk mass, we want to represent this mass as accurately as possible. We cannot track all individual bodies of km-size since their number can exceed $10^{11}$. Instead, we inflated the radius of each particle \citep[see e.g.][]{thebault99}. We found that taking an inflated radius for a 1 km planetesimal, $\rinf/\ab\leq{10}^{-4}$, gave satisfying collision statistics without introducing a bias in encounter velocities.  Collision outcomes are determined based on the velocity-dependent catastrophic disruption criteria of \cite{stewart09} \citep[see][]{collision}. 

\subsection{Gas and dust disk}
The gas disk is assumed to be static with no pressure gradient so that there is no radial drift for solids that are coupled to the gas through aerodynamic drag. While radial drift could be easily incorporated for the planetesimals it will be more important for the small dust component, especially for mass that is in m-sized bodies. Since we have no information about the size distribution of bodies $\le1$ km, we choose to neglect radial drift. For simplicity, the dust surface density is $\propto{r}^{-1}$, so that the mass inside a ring of radial thickness $\Delta{r}$ is constant throughout the disk. Experiments with different density profiles showed no qualitative differences in the outcome of the simulations. Radial migration of planetesimals therefore does not play a major role in our simulations. This may not be true for the small dust component, whose radial drift will be strongly affected by pressure structure in the disk.

\subsection{Dust accretion}
If most collisions between planetesimals are destructive, a large fraction of the solid component of the disk will be in small debris, which can be picked up by remaining planetesimals. This process involves an efficiency factor $\epsilon_\mathrm{d}$ \citep[see][]{collision}. In this letter, we consider the two extreme cases $\epsilon_\mathrm{d}=0$ (no dust accretion) and $\epsilon_\mathrm{d}=1$ (full dust accretion). In view of the results of \cite{housen11}, we switch off dust accretion (even when $\epsilon_\mathrm{d}=1$) whenever the relative velocity between planetesimal and dust exceeds 100 times the escape velocity from the surface of the planetesimal. 

\begin{figure}
\includegraphics[width=\columnwidth]{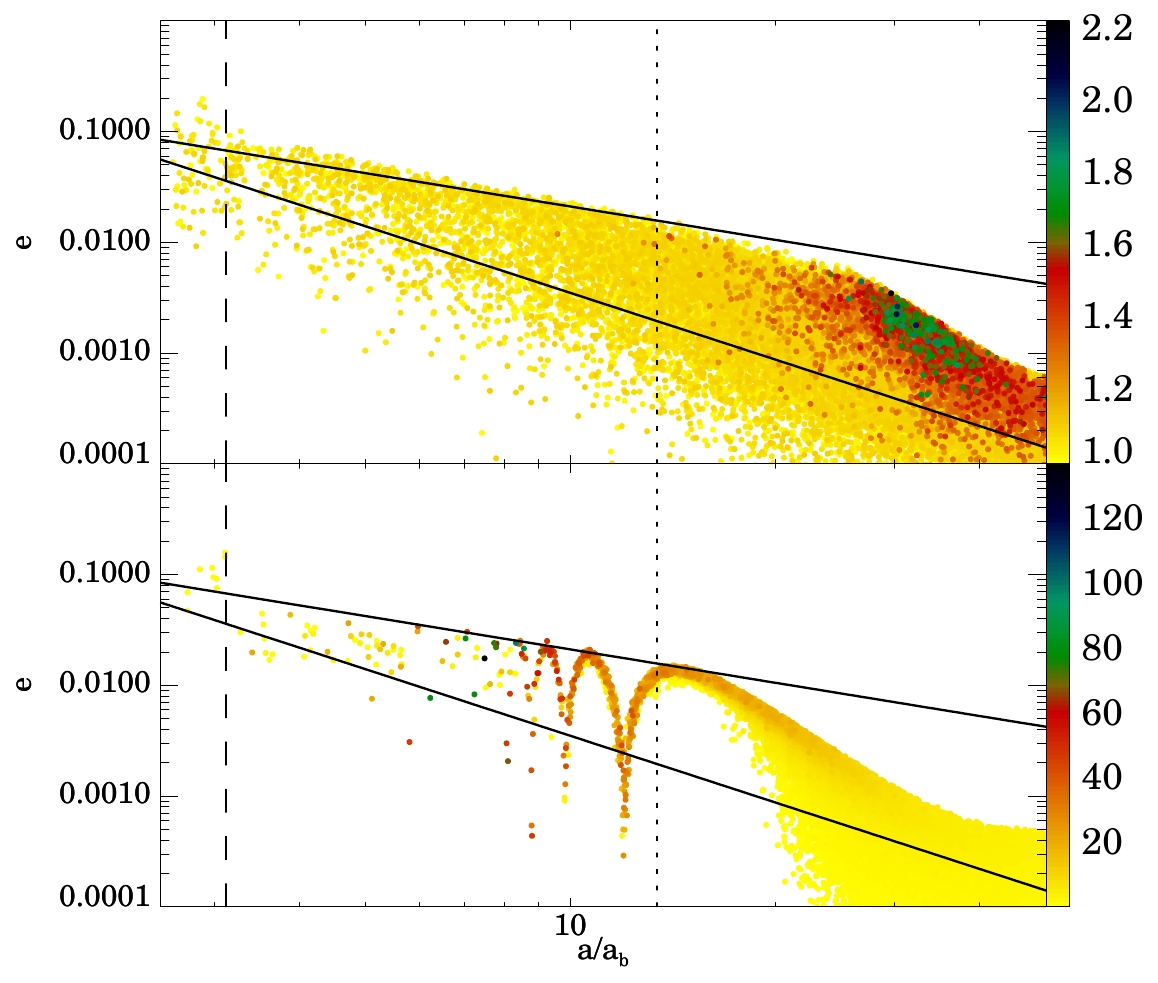}
\caption{Planetesimal eccentricity and semi-major axis for binary parameters of Kepler 16AB. Top panel: no dust accretion,  $t=250,000$ binary orbits. Bottom panel: full dust accretion, $t=80,000$ binary orbits. Colour indicates the size of the planetesimal in km. All darker-coloured particles have accreted mass since the start of the simulation. The vertical dashed line indicates the position of Kepler 16b. The vertical dotted line the inner boundary of the accretion-friendly zone identified by \cite{marzari12}. The solid lines indicate $2\eff$ (steep) and $2\ef$.}
\label{fig_ae}
\end{figure}

\begin{figure}
\includegraphics[width=\columnwidth]{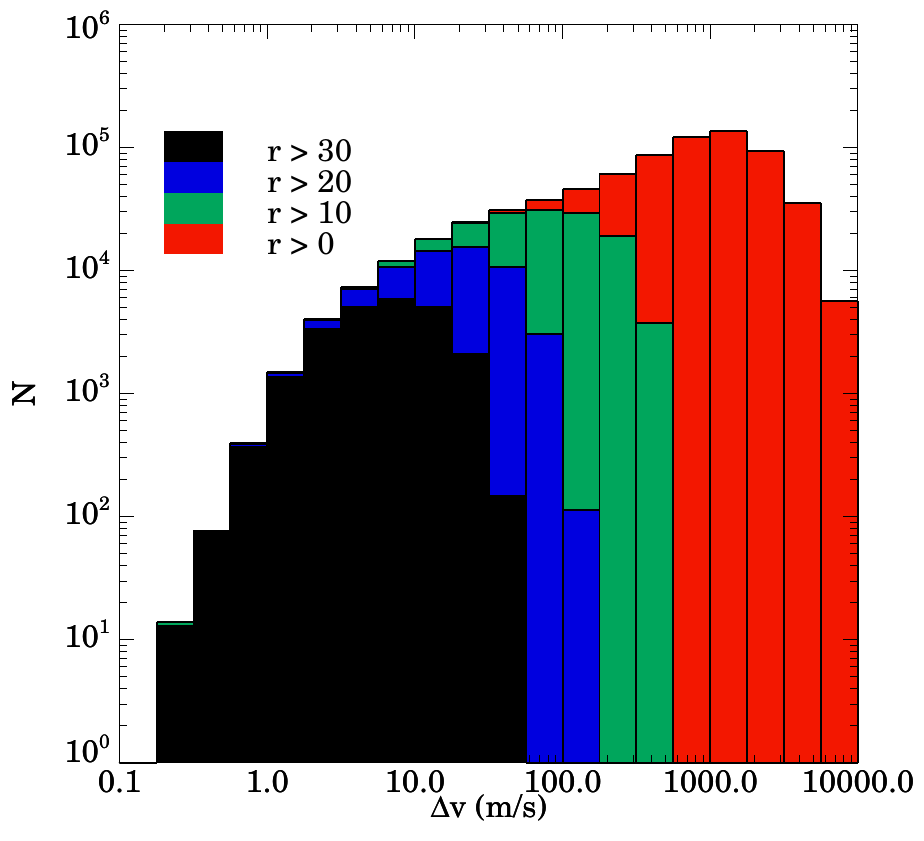}
\caption{Histogram of collision velocities for Kepler 16 scenario without dust accretion, $t=250,000$ binary orbits. Colours indicate different radii, with $r$ in units of $\ab$.}
\label{fig_vc}
\end{figure}

\begin{figure}
\includegraphics[width=\columnwidth]{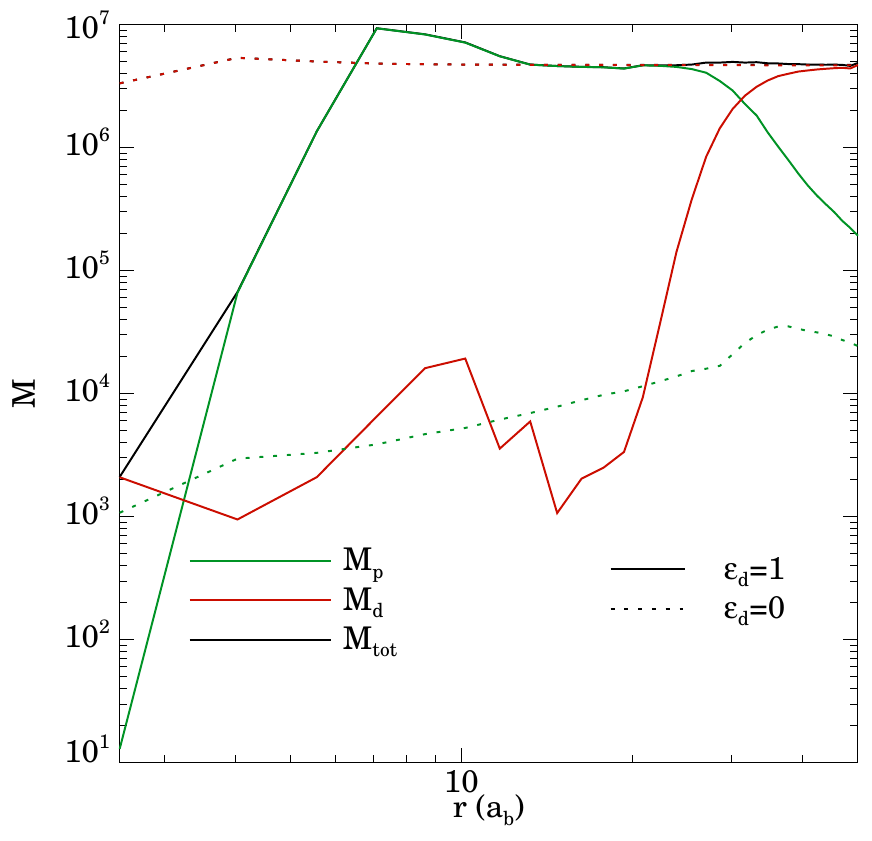}
\caption{Mass distribution in units of 1 km planetesimal for the simulations depicted in Fig. \ref{fig_ae}. Dotted lines: without dust accretion. Solid lines: dust accretion. Green indicates planetesimal mass, red dust mass, and black total mass. Mass is binned radially into 32 bins of equal size, so that initially, $M(r)$ is constant because $\Sigma \propto r^{-1}$.}
\label{fig_mass}
\end{figure}

\section{Results}
\label{secRes}

We focus first on the Kepler 16 system. We take a computational domain for the planetesimal disk $2.5<r/\ab<50$, with a total solid mass of $1.5\cdot{10}^8$ $M_1$, where $M_1$ is the mass of a 1 km planetesimal, and a surface density $\propto{r}^{-1}$. Initially, all the mass is in small dust. Planetesimals form at a rate given by $\epsilon_\mathrm{p}=10^{-5}$ \citep{collision} and have an initial size of 1 km.  The gas disk is assumed to be circular, with a (constant) density of $1.4\cdot{10}^{-9}~\mathrm{g/cm^3}$.

\subsection{No dust accretion}
First we consider a case without dust accretion. Growth can then only occur through collisions. In the top panel of Fig. \ref{fig_ae} we show the eccentricities of planetesimal orbits versus semi-major axis. For the binary parameters of Kepler 16, $\ef$ dominates over $\eff$ throughout the disk. While gas drag acts to slowly force planetesimals onto equilibrium orbits, the continuous introduction of new planetesimals with $e=0$ and the collisional destruction of older planetesimals lead to orbits that are unphased. Collisions between planetesimals on unphased orbits would occur at $v_\mathrm{col}=\ef a\Omega\approx{21}(\ab/a)^{3/2}~\mathrm{km/s}$. This is in good agreement with what is measured from the simulations, see Fig. \ref{fig_vc}. We find that accretion is possible for $a/\ab>20$, or $a>4.4$ AU, in good agreement with results of \cite{meschiari12} and \cite{marzari12}. Even though many collisions around $a/\ab=20$ are still destructive, enough accreting collisions occur to sustain a population of larger planetesimals.  

\subsection{Full dust accretion}
We now turn on dust accretion in full, which is the most favourable case possible for accretion. Dust accretion is quenched for relative velocities exceeding $100{v}_\mathrm{esc}$. The bottom panel of Fig. \ref{fig_ae} shows that in this case it is possible to grow planetesimals further in.  However, even in this most favourable case we find it impossible to grow planetesimals at the location where Kepler 16b resides (vertical dashed line in Fig. \ref{fig_ae}). In regions where accretion is possible, $a/\ab>6$, all the mass ends up in planetesimals, and the maximum planetesimal size that can be reached is only limited by the amount of mass available. This is illustrated in Fig. \ref{fig_mass}, which shows that while for $\epsilon_\mathrm{d}=0$ less than 1\% of the total mass ends up in planetesimals, for $\epsilon_\mathrm{d}=1$ essentially all the mass inside $r/\ab=20$ resides in planetesimals. After $250,000$ binary orbits, the region $r/\ab>20$ has not evolved to the planetesimal-only state yet. Figure \ref{fig_mass} also shows another important effect, namely mass transport towards accretion-friendly regions. The region inside $r/\ab=6$ is depleted, not only in planetesimals because of destructive collisions, but also of small dust. This can happen because planetesimals are not necessarily formed and destroyed at the same radius. If a planetesimal is born at radius $r$ with $e=0$ and its eccentricity is excited to $e=\bar{e}$ its orbit lies between $r(1-\bar{e})$ and $r(1+\bar{e})$. The planetesimal can be destroyed at any radius between these extremes. At the inner edge of the accretion-friendly region, mass gets locked up in large, indestructible planetesimals. Outward mass transport into the accretion-friendly region is therefore a one-way street, which eventually leads to the depletion of the inner disk. This process competes against inward radial dust drift, which we neglect here.
 
\subsection{Other systems}
We now briefly discuss the systems Kepler 34 and Kepler 35. These are both almost equal-mass binaries (see Table \ref{tab1}), which makes $\ef$ small. Planetesimal orbital evolution is predominantly due to short timescale interactions with the binary, whose amplitude falls off rapidly with distance, since $\eff\propto{a}^{-2}$. This makes these systems slightly more accretion friendly than Kepler 16 at small radii. In the case of no dust accretion, we find that planetesimal growth is possible for $a/\ab>12$ for Kepler 34 and for $a/\ab>15$ for Kepler 35. Full dust accretion leads to very similar results: planetesimal accretion is possible slightly further in compared to Kepler 16. However, in situ accretion starting from 1 km planetesimals is still not possible. 

\section{Discussion and conclusions}
\label{secDisc}

We have studied planetesimal collisions in circumbinary gas disks, focusing on the planet-harbouring systems Kepler 16, 34 and 35. We have shown that in addition to secular forcing, planetesimals experience eccentricity forcing on a dynamical timescale, which leads to eccentricity oscillations and orbital crossings that can not be prevented by gas drag. This makes the current location of the planets Kepler 16b, 34b and 35b very hostile for planetesimal accretion. 

We then used a numerical model similar to that of \cite{collision} including planetesimal formation and accretion of small dust. Even in the most favourable case of 100\% efficient dust accretion, we have been unable to grow planetesimals from initially 1 km at the current location of the planets. Since dust accretion is likely to be less than 100\% efficient, for example because not all the small dust will be concentrated in the midplane of the disk, we conclude that in situ planetesimal accretion is difficult for the planets Kepler 16b, 34b and 35b. 

We have made several necessary simplifications to make following the collisional evolution of the planetesimal population tractable. First of all, we have ignored gas dynamics throughout and worked with a static circular gas disk. While the gas disk is likely to become eccentric, especially at small radii, it was shown in \cite{paard08} that unless the gas relaxes towards the forced eccentricity, including gas dynamics makes matters worse for planetesimal accretion. For the fast eccentricity oscillations to be damped by gas drag, the gas disk will have to oscillate in phase with the planetesimals. Full hydrodynamical simulations are necessary to determine whether this is the case. These can also be used to study the effect of the inner truncation of the gas disk, and we will consider  such simulations in future investigations.

We considered the planar case, but a small inclination of the binary plane with respect to the gas disk may promote planetesimal accretion in the circumprimary case \citep{xie09}. However, because of the fast eccentricity oscillations and the resulting orbital crossings it is unclear if this effect can play a role in the circumbinary case. Moreover, it was shown in \cite{fragner11} that including gas dynamics again makes matters worse, even in the misaligned disk case.

A formation mechanism which can leapfrog the problematic km-size range, such as gravitational collapse aided by streaming instabilities \citep{johansen07}, may overcome the problems of planetesimal accretion. It remains to be seen, however, if such a mechanism can operate in close binary systems. Preliminary calculations show that in the current model, we would need to start with planetesimals of at least $10~\mathrm{km}$ in order for in situ accretion of the Kepler circumbinary planets to become possible.

The most straightforward solution is that the three circumbinary planets were assembled further out in an accretion-friendly region, and migrated in towards their current location at a later stage. This can be achieved at a relatively early stage, in the 10-100 km size range, by radial drift due to a pressure gradient in the gas, or at a later stage when the planet is more or less fully grown, by Type I or Type II planetary migration. Whatever the migration mechanism, it is likely that the inner edge of the truncated gas disk will cause migration to stall. We then expect the current location of the planets to be close to the truncation radius of the gas disk.  

\acknowledgments
We thank the referee, H. Perets, for an insightful report. SJP, ZML and CB acknowledge support from STFC.

\end{document}